\documentclass[twocolumn,pra,showpacs,aps]{revtex4-1}
\usepackage{amssymb}
\usepackage{amsmath}
\usepackage{graphicx}
\usepackage{epsfig}

\begin{document}

\title{Topological nodal points in two coupled SSH chains}
\author{C. Li${}^1$, S. Lin${}^1$, G. Zhang${}^2$ and Z. Song${}^1$}
\email{songtc@nankai.edu.cn}
\affiliation{${}^1$School of Physics, Nankai University, Tianjin 300071, China \\
${}^2$College of Physics and Materials Science, Tianjin Normal University,
Tianjin 300387, China}

\begin{abstract}
We study two coupled Su-Schrieffer-Heeger\ (SSH) chains system, which is
shown to contain rich quantum phases associated with topological invariants
protected by symmetries. In the weak coupling region, the system supports
two non-trivial topological insulating phases, characterized by winding
number $\mathcal{N}=\pm 1$, and two types of edge states. The boundary
between the two topological phases arises from two band closing points,
which exhibit topological characteristics in one-dimensional $k$ space. By
mapping Bloch states on a vector field in $k$ space, the band degenerate
points correspond to a pair of kinks of the field, with opposite topological
charges. Two topological nodal points move and merge as the inter-chain
coupling strength varies. This topological invariant is protected by the
translational and inversion symmetries, rather than the antiunitary
operation. Furthermore, we find that when a pair of nodal points is created,
a second order quantum phase transition (QPT) occurs, associating with a gap
closing and spontaneously symmetry breaking. This simple model demonstrates
several central concepts in the field of quantum materials and provides a
theoretical connection between them.
\end{abstract}

\maketitle



\section{Introduction}

Topological gapless systems have emerged as a new frontier in the field of
quantum materials \cite%
{Wan,Yang,Burkov,Xu,Kim,Weng,Huang,Young,Wang,Wang1,Hou,Sama,Liu,Neupane,SXu,Lv,Lu}%
. As a joint of two quantum phases, topological gapless systems have band
structures with band-touching points in the momentum space, where these kind
of nodal points appear as topological defects of an auxiliary vector field.
Then these points are\ unremovable due to the symmetry protection, until a
pair of them meet together and annihilate. In general, the studied systems
are usually $2$-D and $3$-D with the broken time-reversal symmetry. The
central goal of this work is to understand the physics of topological
gapless system from the point of view of the $1$-D model. A particular
advantage in working within the $1$-D system is that all the parameters of
this model can be easily accessed within the existing technology of
cold-atomic experiments \cite{Clay,Ueda,Jo}. On the other hand, we find that
unlike the topological gapless systems in $2$-D, the magnetic flux is not
necessary for the quasi $1$-D system.

In this work, we systematically study two coupled Su-Schrieffer-Heeger\
(SSH) chains system. Comparing to SSH chain, this ladder system contains
rich quantum phases associated with topological invariants protected by
symmetries. In the weak coupling region, the system supports two non-trivial
topological insulating phases, characterized by winding number $\mathcal{N}%
=\pm 1$, and two types of edge states for the open ladder. We focus on the
boundary between these two quantum phases. We find that the boundary phase
arises from two band closing points, which exhibit topological
characteristics in the one-dimensional $k$ space. We investigate the
topological feature of the boundary by mapping Bloch states on a vector
field in $k$ space and find that, the band degenerate points correspond to a
pair of kinks of the field, with opposite topological charges. As the
inter-chain coupling strength varies, two topological nodal points move and
merge in $k$ space. In contrast to the case of $2$-D\ semimetal phase, the
topological invariant is protected by the translational and inversion
symmetries, rather than the antiunitary operation.

In conventional QPTs, quantum phases are differentiated by the symmetry \cite%
{Sachdev}, while topological matters are classified according to topological
invariants \cite{Kane,Zhang}. The possible connection between two regimes is
studied in a class of quantum spin models \cite{Gang}. In this work, we find
that when a pair of nodal points is created, a second order quantum phase
transition (QPT) occurs, associating with a gap closing and spontaneously
translational and inversion symmetry breaking. This simple model provides an
alterative example to demonstrate the coexistence two types of QPTs,
revealing the connection between them.

The remainder of this paper is organized as follows. In Sec. \ref{Model and
phase diagram}, we present a two coupled-SSH chain model and the quantum
phase diagram, characterized by winding number and edge states. Sec. \ref%
{Topological feature of zero points} reveals the topological feature of
nodal points. Sec. \ref{Perturbations} demonstrates the behavior of
topological gapless system in the presence of several types of
perturbations. Sec. \ref{Symmetry protection of kinks} devotes to the
symmetries that protect the topological invariants. Finally, we present a
summary and discussion in Sec. \ref{Summary}.

\begin{figure}[tbp]
\includegraphics[ bb=22 346 372 785, width=0.45\textwidth, clip]{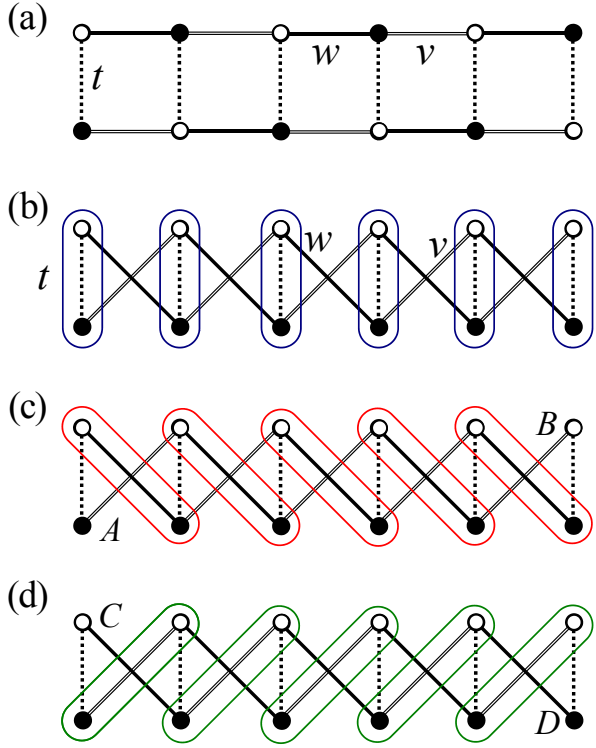}
\caption{(Color online) Schematics of the two coupled SSH chains and the
formation of edge states. The system consists of two sublattices $A$ and $B$,
indicated by filled and empty circles, respectively (a) Hopping amplitudes
along each chain are staggered by $v$ (double line) and $w$ (single line).
The interchain hopping amplitude is $t$ (dotted line). (b) Deformation of
the system without changing the lattice structure. In this geometry a unit
cell (circled by the blue line) contains two sites. In the case of $t\gg v$,$w>0$, the interchain dimerization (circled by the blue line) results in
topologically trivial insulating phase, since there is no edge state. (c) In
the limit $w\gg v$, $t>0$, edge states form at sites A and B. (d) Edge
states form at sites C and D in the case $v\gg w$, $t>0$.} \label{fig1}
\end{figure}

\section{Model and phase diagram}

\label{Model and phase diagram}

The SSH model \cite{SSH} has served as a paradigmatic example of the $1$-D
system supporting topological character \cite{Zak}. It has an extremely
simple form but well manifests the typical feature of topological insulating
phase, and the transition between non-trivial and trivial topological
phases, associated with the number of zero energy edge states as the
topological invariant \cite{Asboth}.

We consider a two coupled SSH chains system, which has a two-leg ladder
structure. Fig. \ref{fig1}(a) sketches the geometry of the system, in which
the hopping amplitudes in each leg is staggered and the rung indicates the
interleg hopping. Such a ladder system is a bipartite lattice system,
consisting of two sub-lattices $A$ and $B$. We write down the Hamiltonian
for the system in a simple form%
\begin{equation}
H=\sum_{j=1}^{N}(wa_{j+1}^{\dagger }b_{j}+va_{j}^{\dagger
}b_{j+1}+ta_{j}^{\dagger }b_{j})+\mathrm{H.c.,}  \label{H}
\end{equation}%
where $a_{l}^{\dag }$ and $b_{l}^{\dag }$ are the creation operators of
fermion at $l$th site of sub-lattice $A$ and $B$, respectively. For a
periodic boundary condition, which just like a ring, we take $%
a_{N+1}^{\dagger }=a_{1}^{\dagger }$ and $b_{N+1}^{\dagger }=b_{1}^{\dagger }
$, while for an open one, we take $a_{N+1}^{\dagger }=b_{N+1}^{\dagger }=0$.
The ring situation can be a description for the bulk part of the system,
i.e., the long central part of the open ladder. Taking the transformation%
\begin{equation}
\left\{
\begin{array}{c}
a_{k}=\frac{1}{\sqrt{N}}\sum_{j}e^{ikj}a_{j} \\
b_{k}=\frac{1}{\sqrt{N}}\sum_{j}e^{ikj}b_{j}%
\end{array}%
\right. ,  \label{Fourier}
\end{equation}%
we have
\begin{equation}
H=\sum_{k}H_{k}=\sum_{k}(a_{k}^{\dagger },b_{k}^{\dagger })h_{k}\left(
\begin{array}{c}
a_{k} \\
b_{k}%
\end{array}%
\right) ,
\end{equation}%
where%
\begin{equation}
h_{k}=\left(
\begin{array}{cc}
0 & we^{ik}+ve^{-ik}+t \\
we^{-ik}+ve^{ik}+t & 0%
\end{array}%
\right) ,
\end{equation}%
and the wave vector $k=\pi (2n-N)/N$, $(n=0,1,...,N-1)$. The Hamiltonian $H$
can be easily diagonalized since $[H_{k},H_{k^{\prime }}]=0$. Therefore, the
phase boundary can be obtained by the zero points of the spectrum

\begin{equation}
\varepsilon _{k}=\pm \sqrt{\left\vert we^{-ik}+ve^{ik}+t\right\vert ^{2}}.
\label{energy band}
\end{equation}%
From $\varepsilon _{k}=0$, we have equations%
\begin{equation}
\left\{
\begin{array}{l}
\left( w+v\right) \cos k_{c}+t=0, \\
\left( w-v\right) \sin k_{c}=0,%
\end{array}%
\right.
\end{equation}%
which determine the phase boundary and the position of band degeneracy point
$k_{c}$. There are two types of boundaries, depending on the value of $k_{c}$%
.\ For $\sin k_{c}\neq 0$\ we have $w=v$ and $\cos k_{c}=-t/(2v)$. Then the
first type of the boundary is%
\begin{equation}
w=v\text{, }\left\vert 2v/t\right\vert >1,  \label{boundary1}
\end{equation}%
which corresponds to two band degeneracy points $\pm \left\vert
k_{c}\right\vert $. On the other hand, for $\sin k_{c}=0$, we have $\cos
k_{c}=-t/(w+v)$, it results in the second type of the boundary
\begin{equation}
\left\vert \left( w+v\right) /t\right\vert =1,  \label{boundary2}
\end{equation}%
which corresponds to a single band degeneracy point $\pi $\ or $-\pi $. We
note that there are two triplet points at $w/t=v/t=\pm 1/2$, which are
joints\ connecting two types of boundary. We will see that such two types of
boundaries correspond to topologically trivial and non-trivial states.

Now we focus on the quantum phases separated by the obtained boundaries.
Matrix $h_{k}$\ is the core of the Hamiltonian, containing all the
information of the system. We rewrite $h_{k}$\ as the form%
\begin{equation}
h_{k}=\mathbf{d(}k\mathbf{)}\cdot \mathbf{\sigma }
\end{equation}%
with a $3$-D vector%
\begin{equation}
\left\{
\begin{array}{l}
d_{x}\mathbf{(}k\mathbf{)}\mathbf{=}\left( w+v\right) \cos k+t, \\
d_{y}\mathbf{(}k\mathbf{)}\mathbf{=}\left( v-w\right) \sin k, \\
d_{z}\mathbf{(}k\mathbf{)}\mathbf{=}0,%
\end{array}%
\right.
\end{equation}%
where $\mathbf{\sigma }$ represents 3-D\ Pauli matrix. The topology of the
energy bands in each areas can be characterized by the loop of the curve
\cite{Gang}%
\begin{equation}
\left\{
\begin{array}{l}
x\mathbf{=}\left( w+v\right) \cos k+t, \\
y\mathbf{=}\left( v-w\right) \sin k,%
\end{array}%
\right.  \label{loop}
\end{equation}%
in the auxiliary space $(x,y)$. The feature of the quantum phase is
characterized by the topology of the loop. The winding number of the loop
around the origin of $xy$ plane is defined as

\begin{equation}
\mathcal{N}=\frac{1}{2\pi }\int\nolimits_{c}\frac{1}{r^{2}}\left( x\mathrm{d}%
y-y\mathrm{d}x\right) ,  \label{winding number}
\end{equation}%
where $r^{2}=x^{2}+y^{2}$.\ A straightforward derivation from above
definition yields%
\begin{equation}
\mathcal{N}=\left\{
\begin{array}{cc}
0, & \left\vert t\right\vert >\left\vert w+v\right\vert  \\
\mathrm{sgn}(v^{2}-w^{2}), & \text{otherwise}%
\end{array}%
\right. ,  \label{N}
\end{equation}%
where \textrm{sgn}(.) denotes the sign function.\ It shows that there are
three phases with $\mathcal{N}=0,$\ $\pm 1$, respectively. Actually, the
conclusion of Eq. (\ref{N}) can be obtained directly from the geometry of
the loop represented by the parametric equations (\ref{loop}). Obviously,
the loop is an ellipse with center located at $(t,0)$ in the $xy$ plane,
with the length of semiaxis along the $x$ (or $y$) axis is $\left\vert
w+v\right\vert $ (or $\left\vert w-v\right\vert $) (see Fig. \ref{fig5}).
When $\left\vert t\right\vert >\left\vert w+v\right\vert $, the origin is
out of the ellipse, resulting zero winding number. Otherwise, the winding
number is $\pm 1$. On the other hand, as $k$\ increases from $-\pi $\ to $%
\pi $, the rotating direction of the loop depends on the sign of the ratio $%
[(\partial x/\partial k)/(\partial y/\partial k)]_{k=\pi /4}$ $=-\left(
w+v\right) /\left( v-w\right) $, or the sign of $\left( v^{2}-w^{2}\right) $.

\begin{figure}[tbp]
\includegraphics[ bb=36 330 468 744, width=0.45\textwidth, clip]{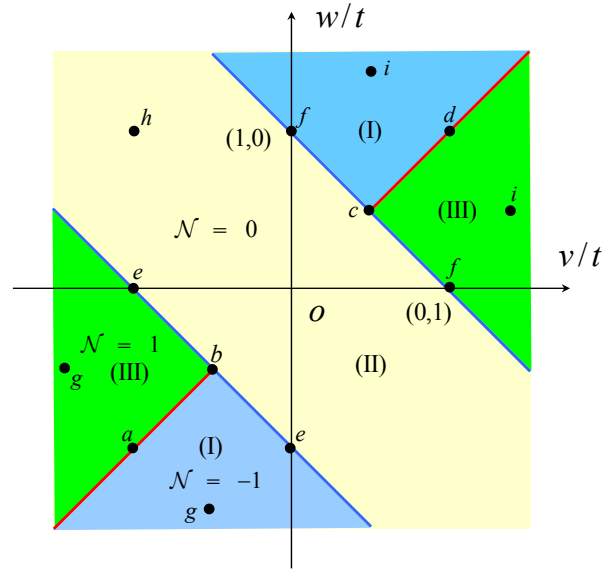}
\caption{(Color online) Phase diagram of the coupled SSH chains system on the $vw$
plane (in unit of t). The green line indicates the boundary separated the
phases with Chern number $c=\pm1$. The blue line indicates the boundary separated
the phases with Chern number $\vert c\vert=1$ and $0$.} \label{fig2}
\end{figure}

\begin{figure*}[tbp]
\includegraphics[ bb=10 251 331 520, width=0.23\textwidth, clip]{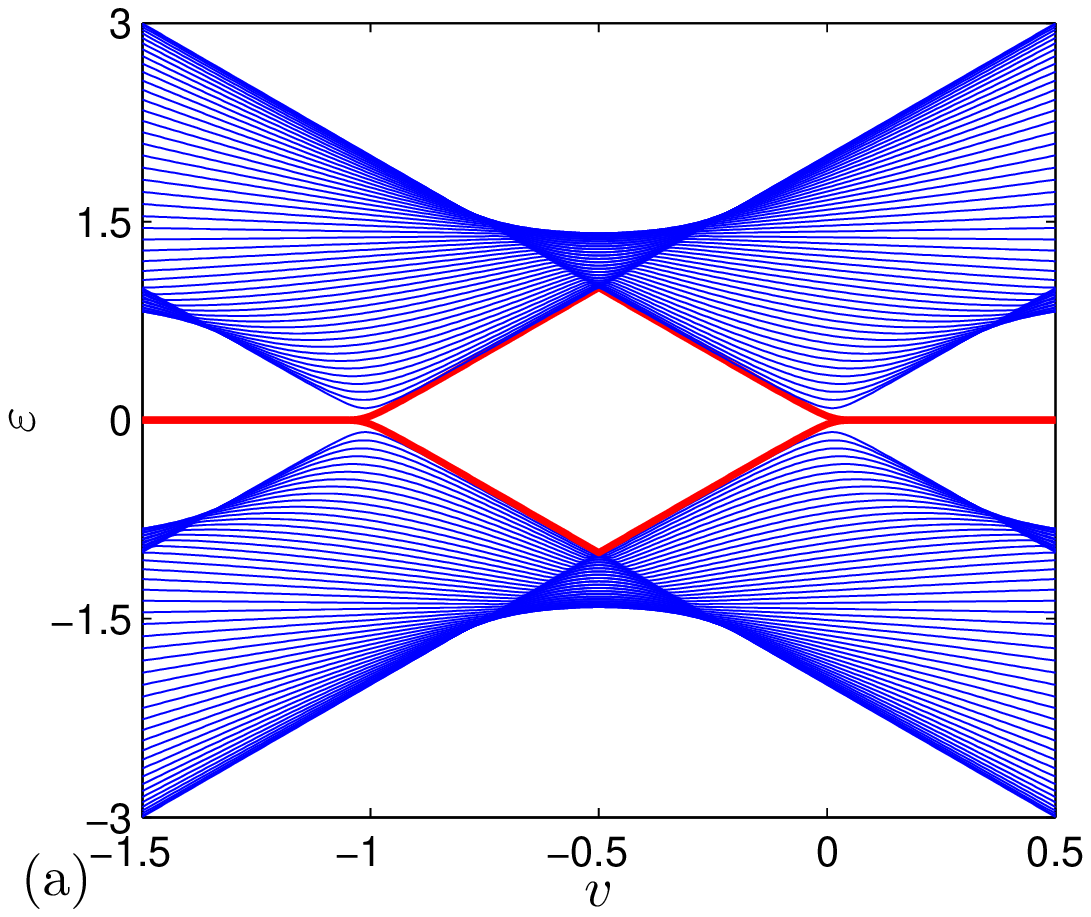} %
\includegraphics[ bb=10 251 331 520, width=0.23\textwidth, clip]{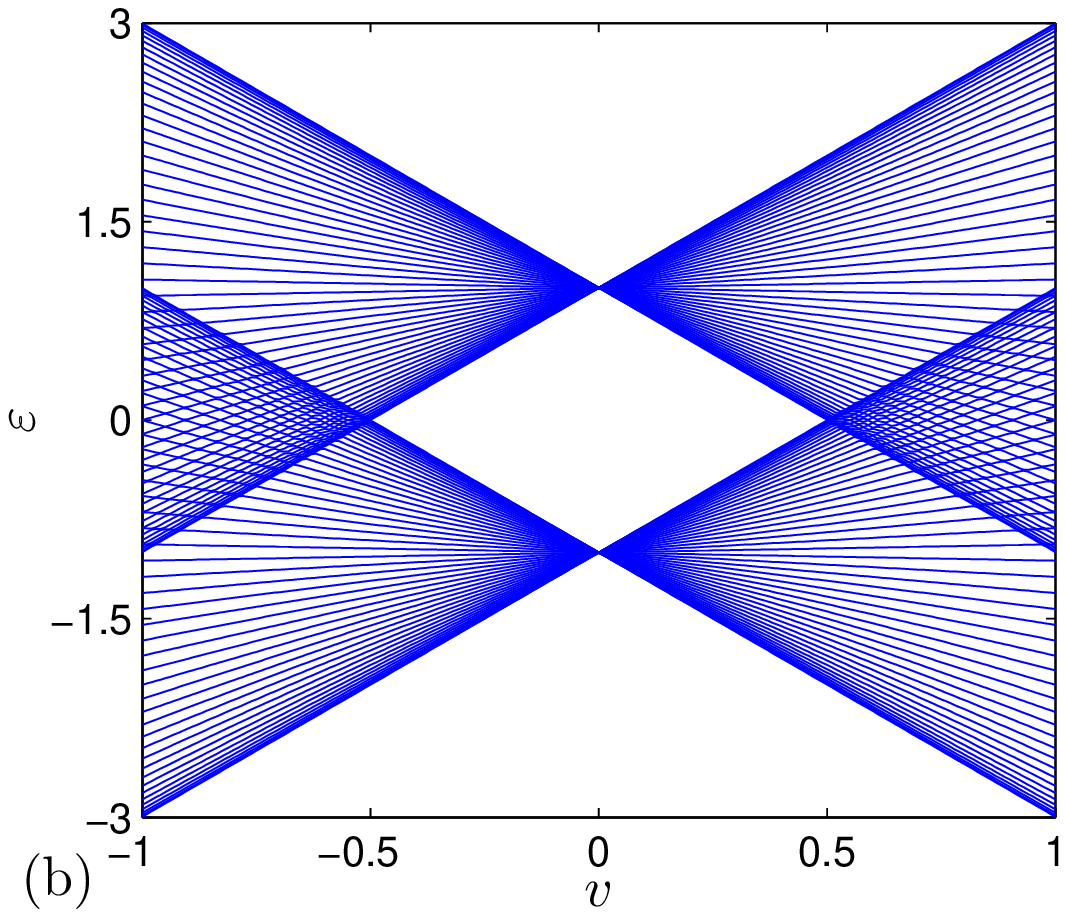} %
\includegraphics[ bb=10 251 331 520, width=0.23\textwidth, clip]{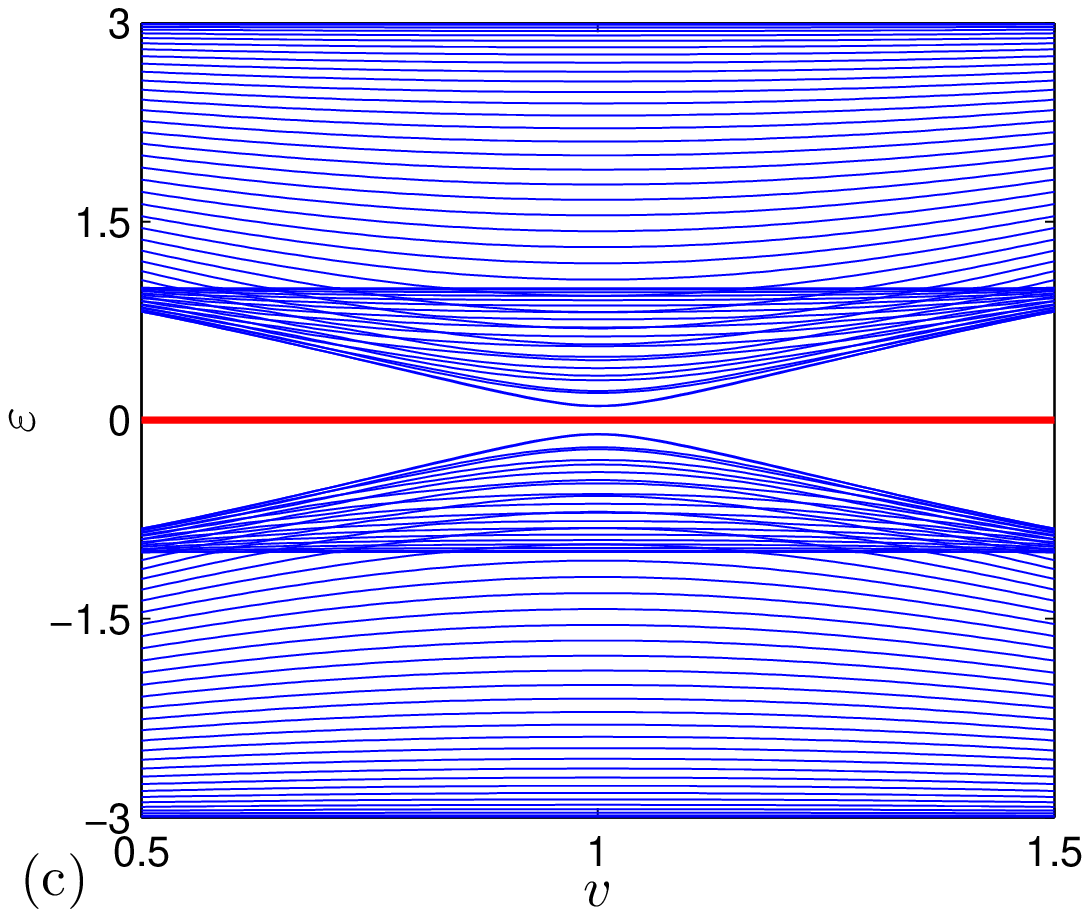} %
\includegraphics[ bb=10 251 331 520, width=0.23\textwidth, clip]{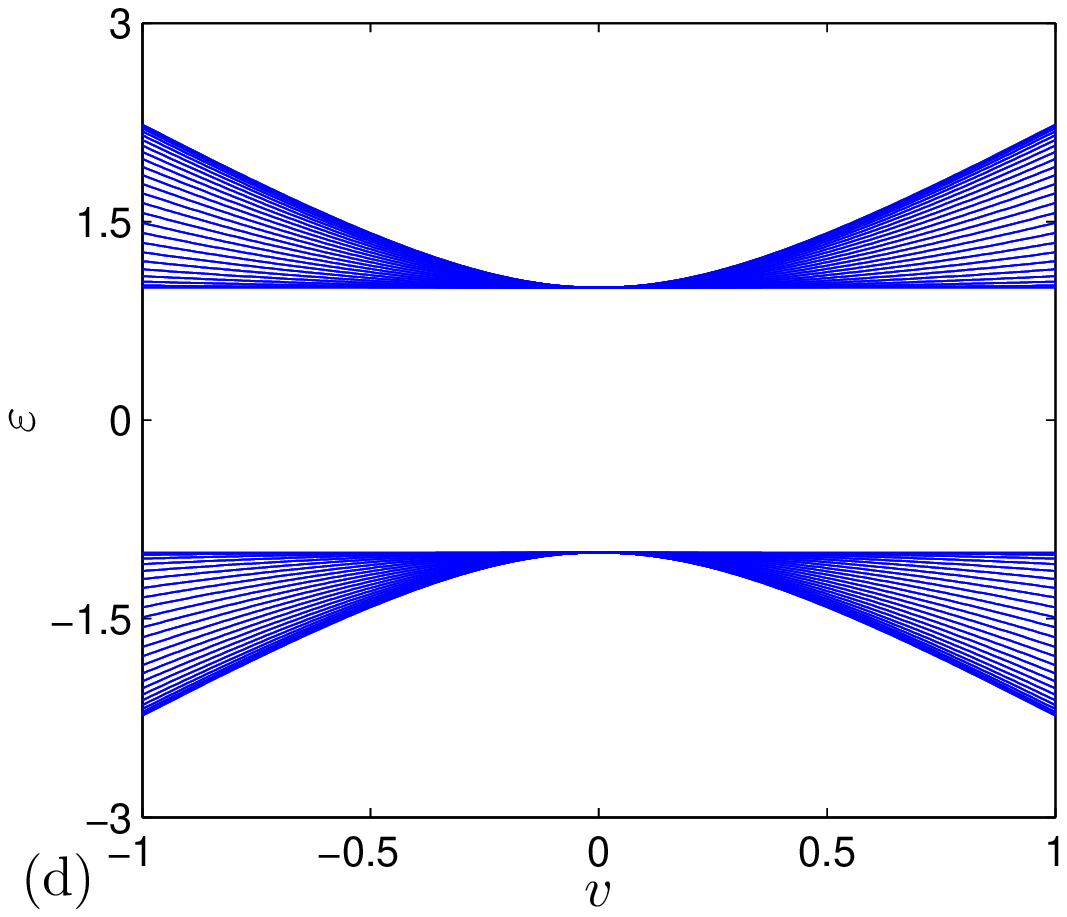}
\caption{(Color online) Energy spectra for the Hamiltonian in Eq. (\protect
\ref{H}) as a function of $v/t$\ on a $2\times N$ ladder with the open
boundary condition, obtained by exact diagonalization. The parameters obey
the relations (a) $v/t-w/t=-1$, (b) $v/t-w/t=0$\textbf{,}\ (c) $v/t+w/t=2$,
and (d) $v/t+w/t=0$. We see that energy gaps open in all three regions and
the zero modes appear in the regions of (I) and (III) in the case of (a).
The energy gap opens only in the region of (II) in the case of (b). The zero
points in the regions of (I) and (III) are consisted of crossing points. In
(c), there are always zero modes appear and the gap nearly close when arrive
the point $v=w=1$. In (d), no zero modes appear all the time.}
\label{fig3}
\end{figure*}

We plot the phase diagram in Fig. \ref{fig2}\textbf{.} According to the
bulk-edge correspondence \cite{Kane,Zhang}, there should be gapless edge
states when the bulk states are topologically nontrivial. Therefore, the
quasi-zero modes should appear for finite open ladder when the parameters
are taken in the region (I) and (III), but be absent in the region (II). To
demonstrate this point, we plot the energy levels as functions of $v/t$\
along the lines (a) $v/t-w/t=-1$, (b) $v/t-w/t=0$,\ (c) $v/t+w/t=2$, and (d)
$v/t+w/t=0$,\ in the phase diagram, respectively. The plot in Fig. \ref{fig3}
are in agreement with the analysis, manifesting topological zero-energy
modes for non-trivial areas.

\begin{figure*}[tbp]
\includegraphics[ bb=3 232 344 511, width=0.3\textwidth, clip]{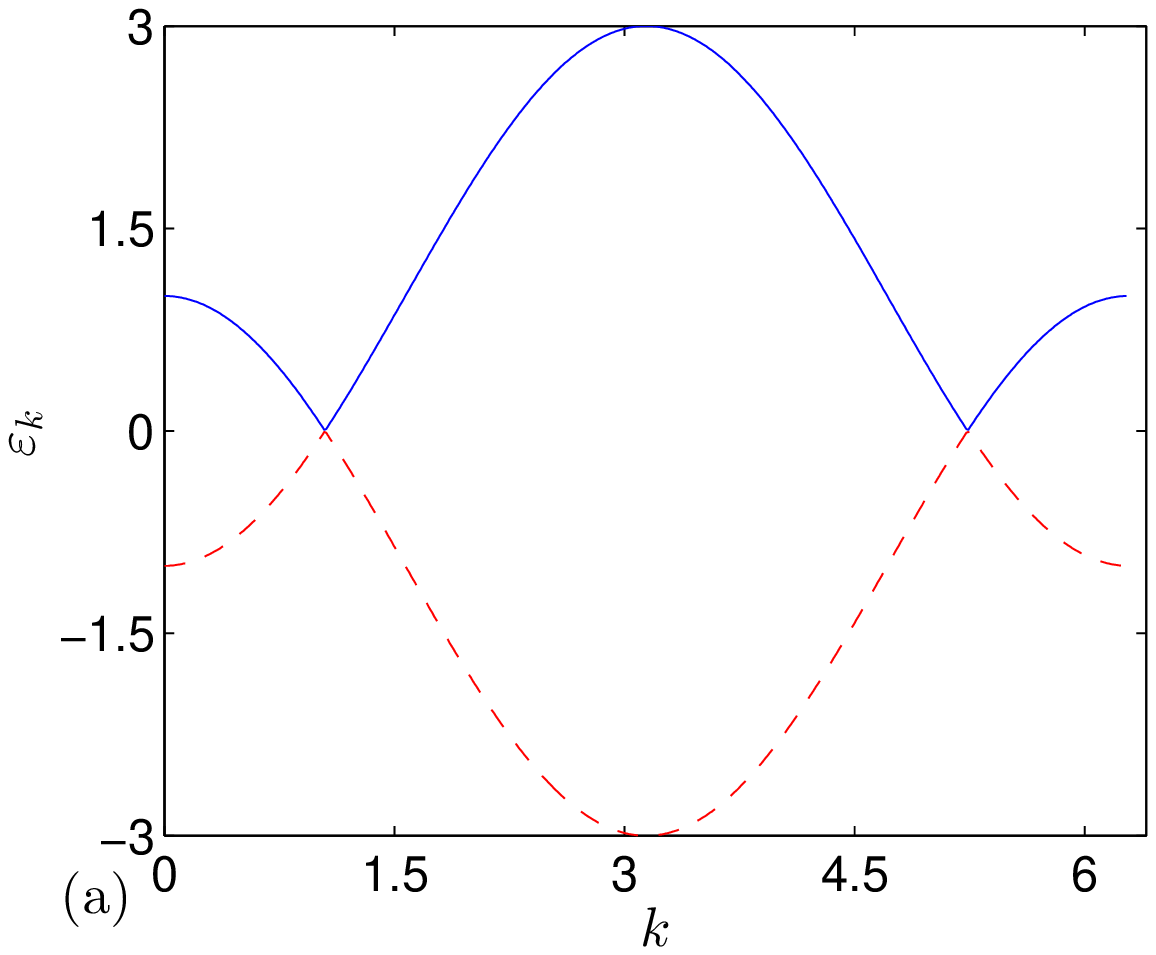} %
\includegraphics[ bb=3 232 344 511, width=0.3\textwidth, clip]{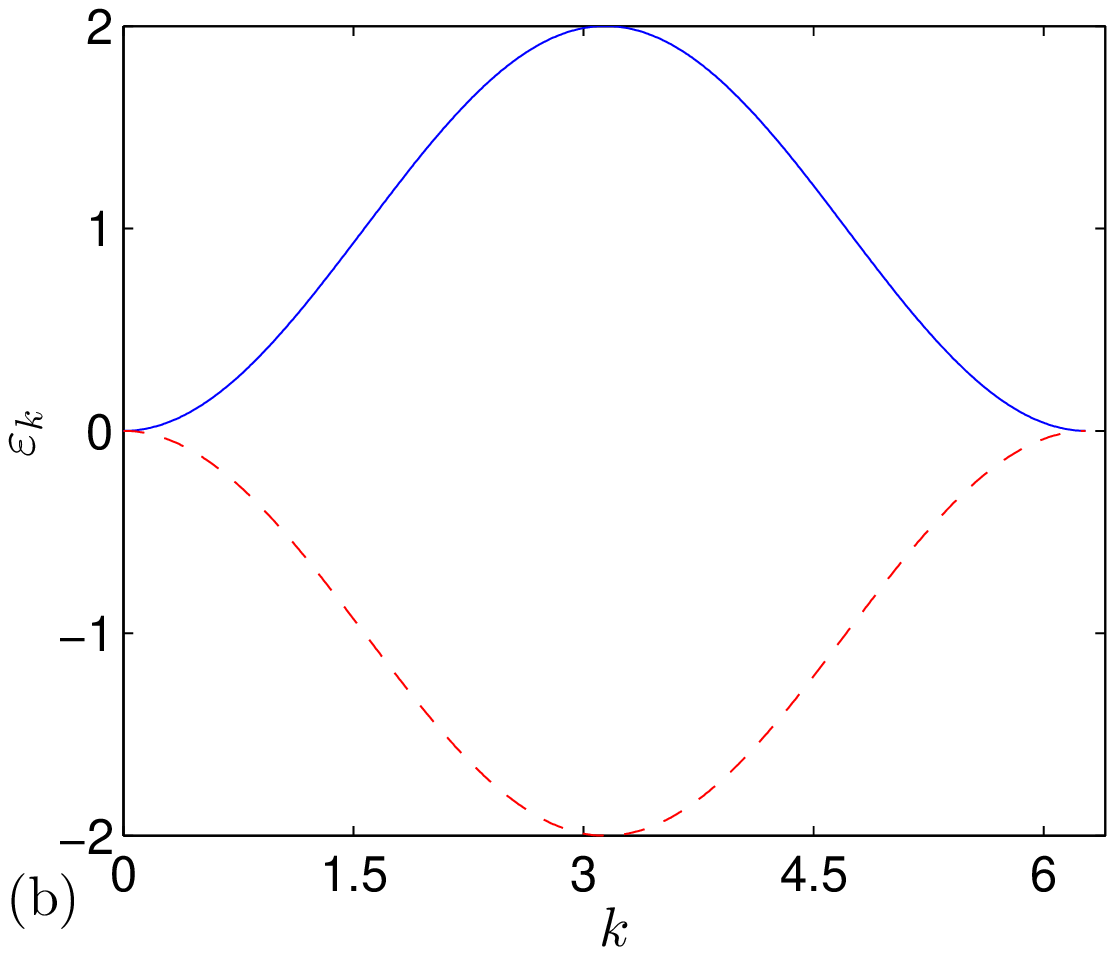} %
\includegraphics[ bb=3 232 344 511, width=0.3\textwidth, clip]{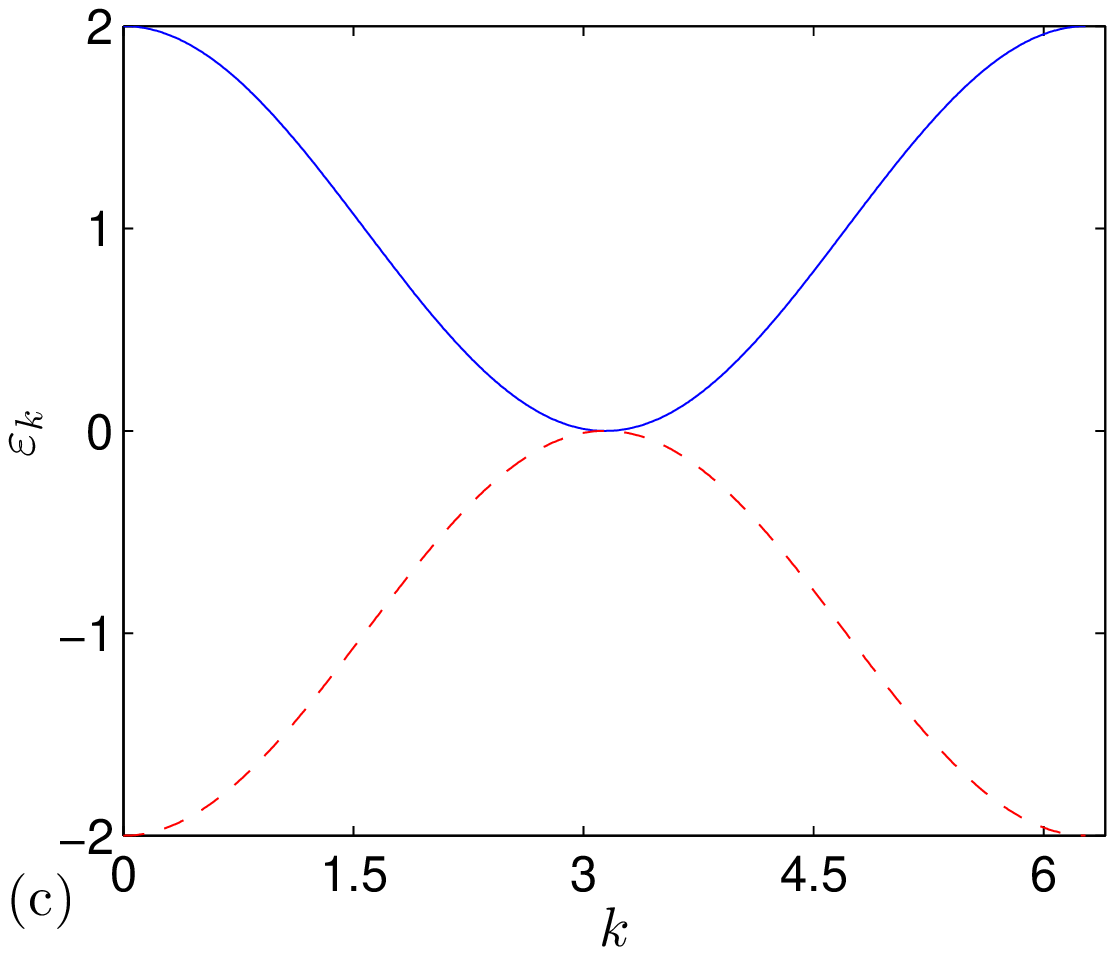} %
\includegraphics[ bb=3 232 344 511, width=0.3\textwidth, clip]{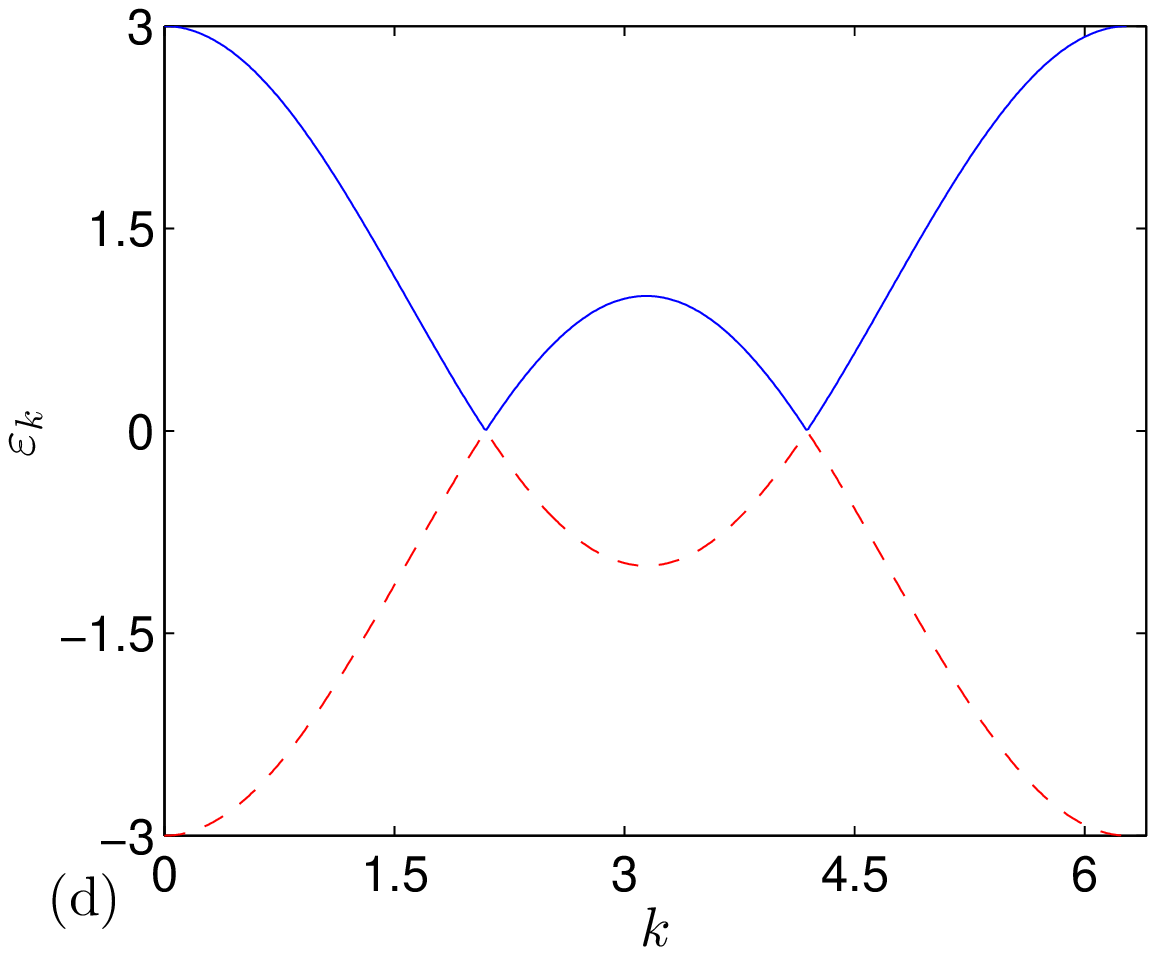} %
\includegraphics[ bb=3 232 344 511, width=0.3\textwidth, clip]{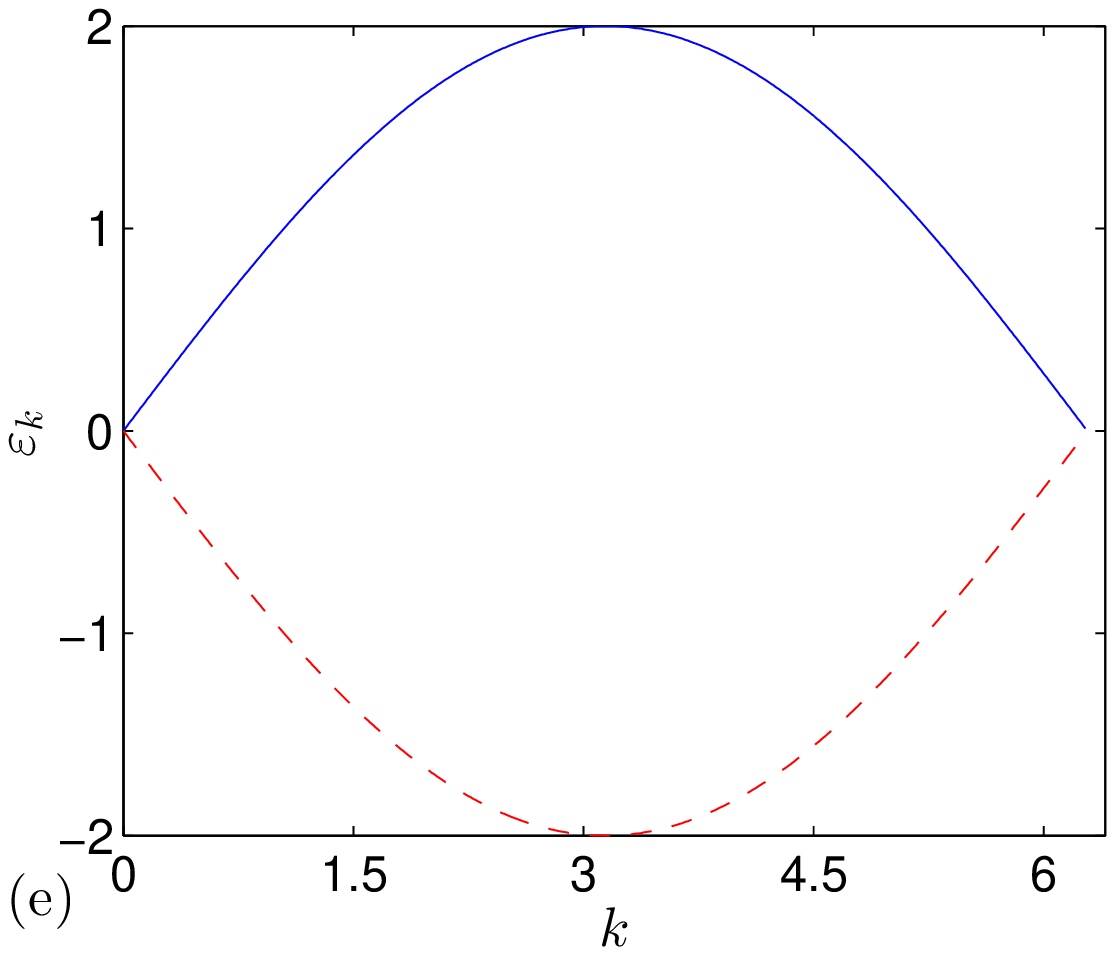} %
\includegraphics[ bb=3 232 344 511, width=0.3\textwidth, clip]{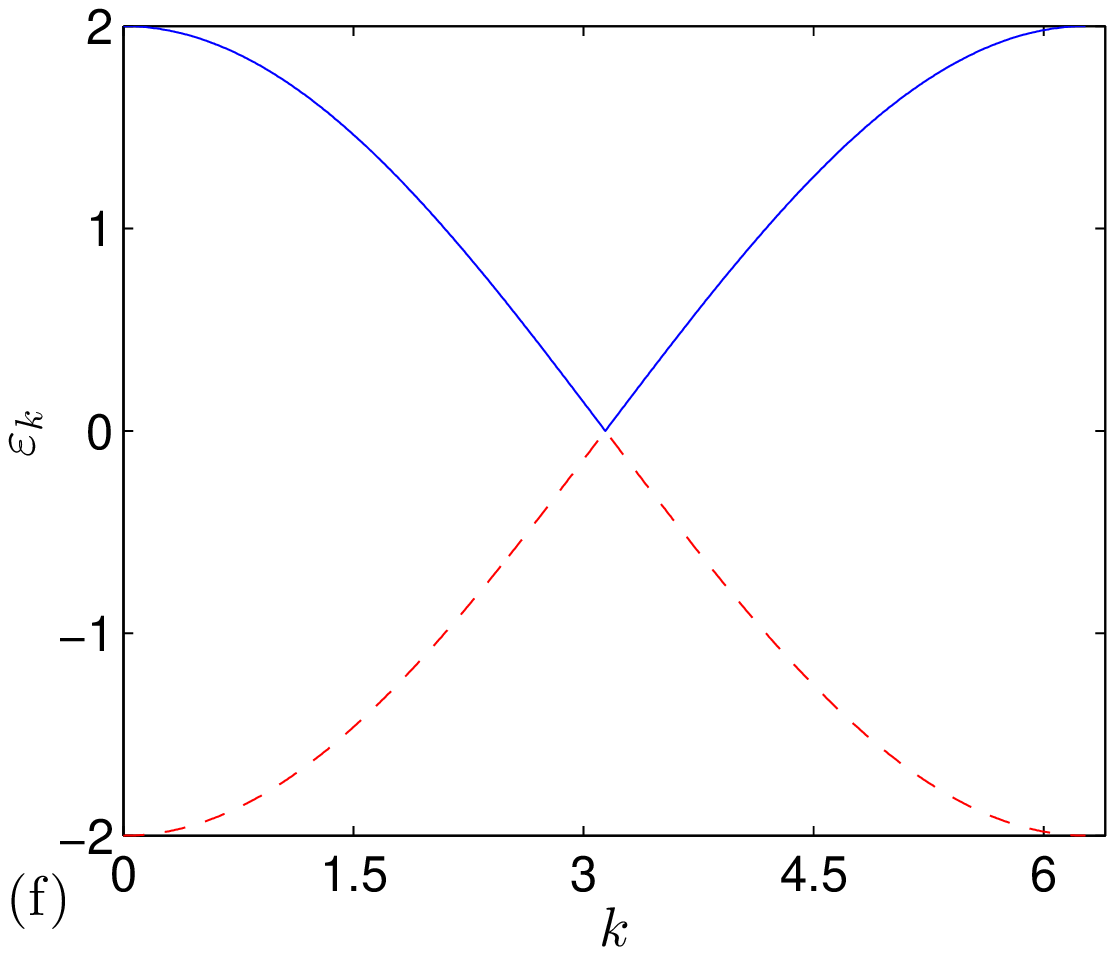} %
\includegraphics[ bb=3 232 344 511, width=0.3\textwidth, clip]{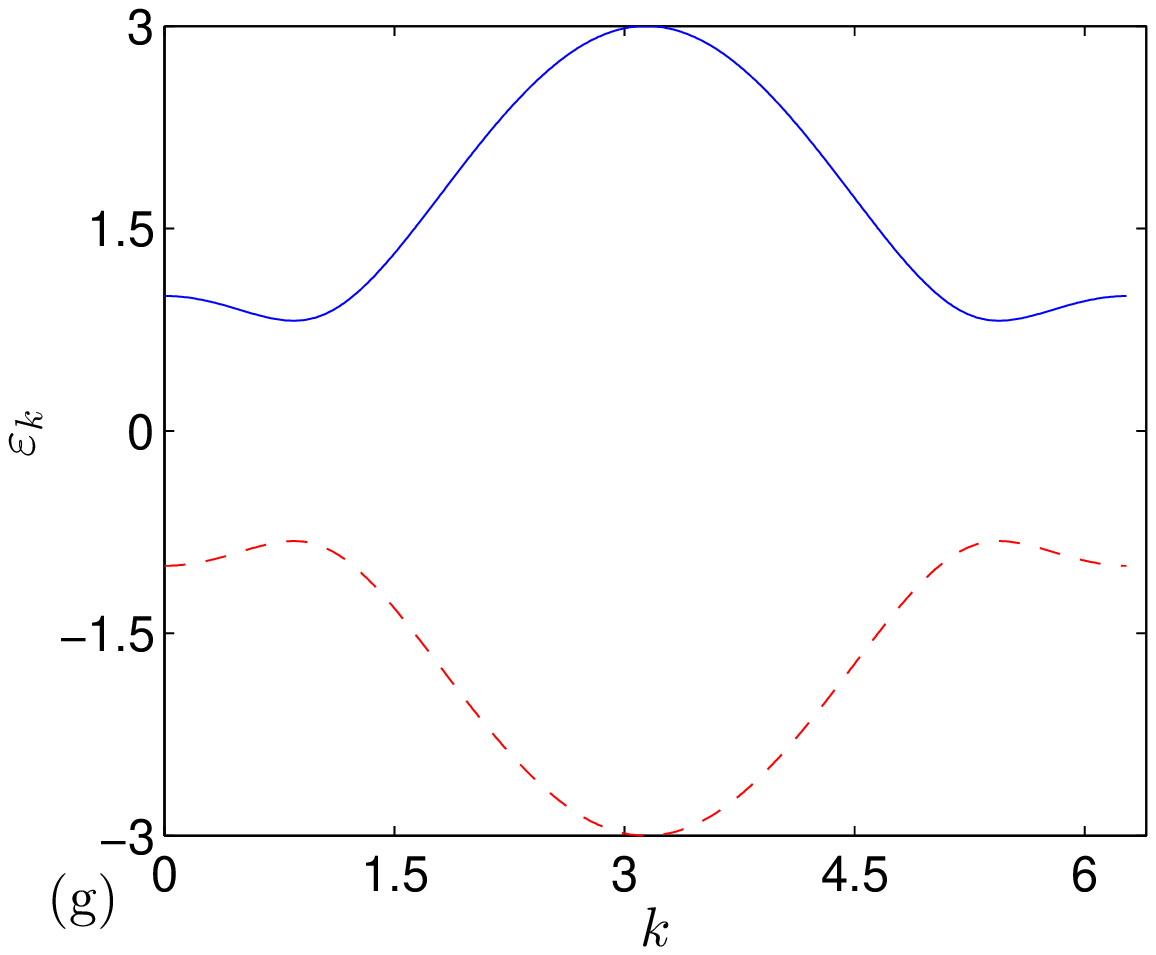} %
\includegraphics[ bb=3 232 344 511, width=0.3\textwidth, clip]{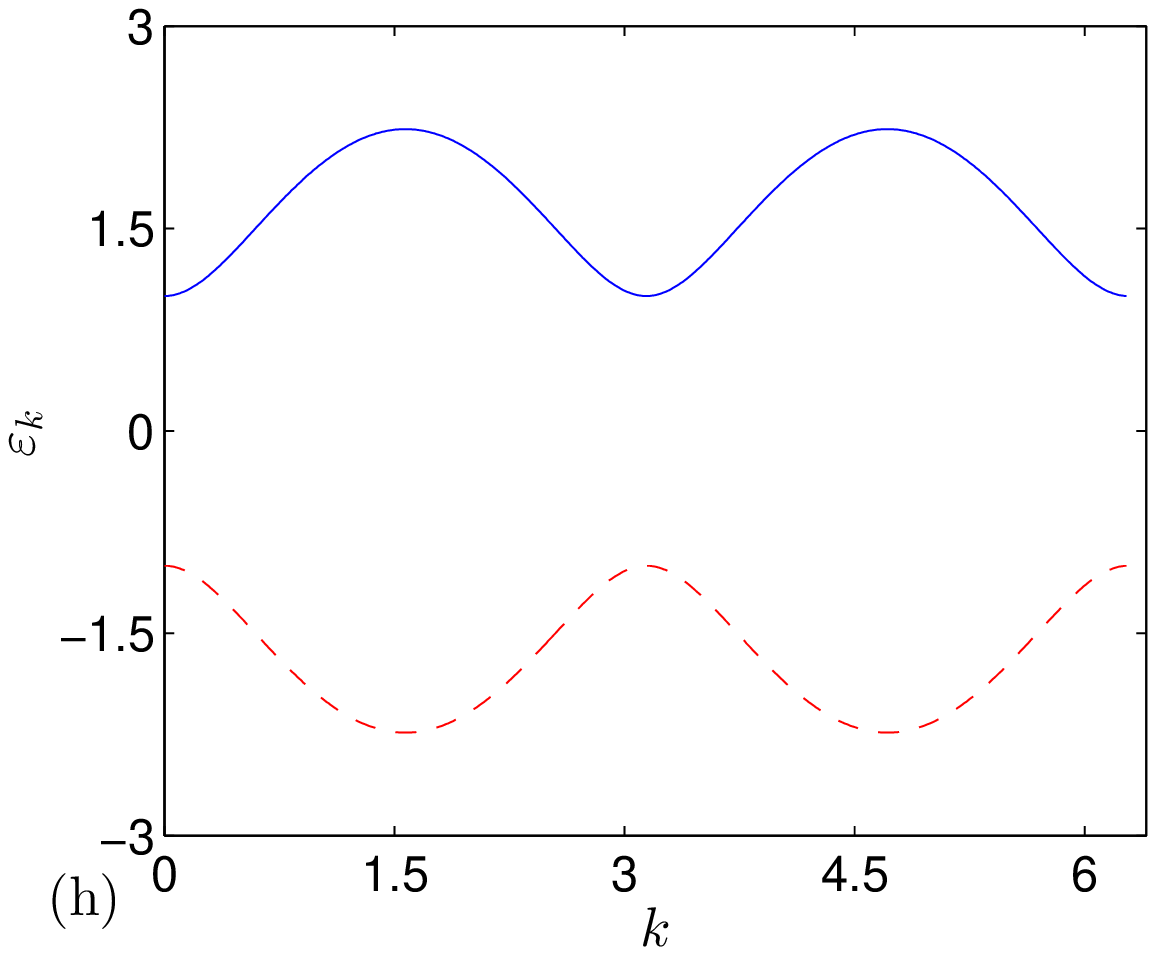} %
\includegraphics[ bb=3 232 344 511, width=0.3\textwidth, clip]{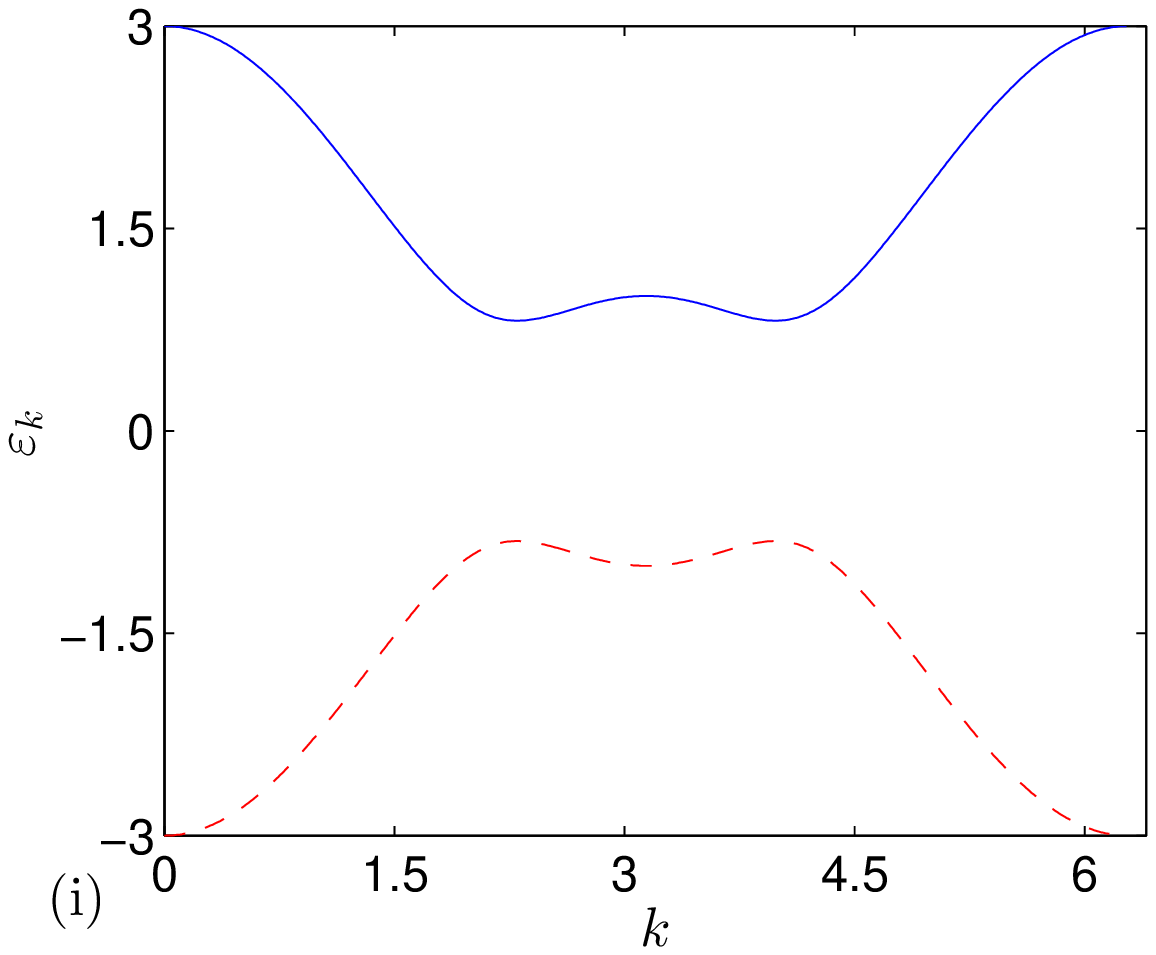}
\caption{(Color online) Plots of the structure of the energy band described
by Eq. (\protect\ref{energy band}) at some points fixed in the Fig. \protect
\ref{fig2}. $t=1$ for all figures. Graphs (a)-(f) just show the energy band
situation in the phase boundaries $v=w$ and $v+w=\pm 1$. Meanwhile (g)-(i)
display what energy band situation is in different phase areas. One can see
that some symmetric points in Fig. \protect\ref{fig2} have the same band
structure.}
\label{fig4}
\end{figure*}

\begin{figure}[tbp]
\includegraphics[ bb=60 455 541 780, width=0.45\textwidth, clip]{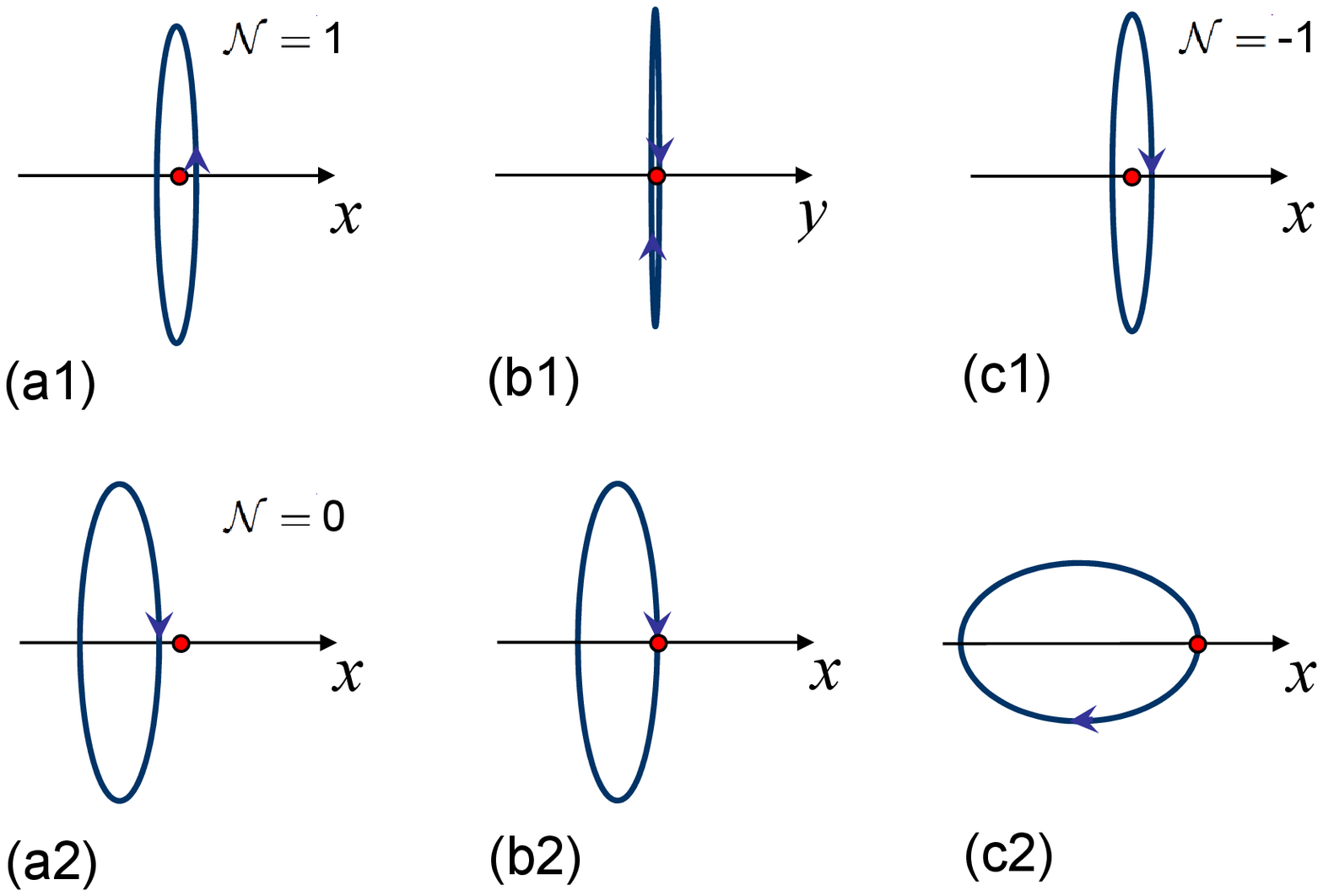}
\caption{(Color online) Schematic illustration of two types of boundaries by
the geometry of graphs in the auxiliary $xy$ plane. Red filled circle
indicates the origin. Winding numbers and graphs of non-trivial topological
phases with (a1) $v^{2}-w^{2}>0$, (c1) $v^{2}-w^{2}<0$. The non-trivial
topological boundary with (b1) $w=v$, always corresponds to the graph of a
segment. (a2) Graph and winding number of a trivial topological insulating
phase. (b2) and (c2) Graphs of trivial topological boundary. They correspond
to ellipses with various shapes, but passing through the origin. We see that
two types of boundaries correspond to two completely different types of graphs.}
\label{fig5}
\end{figure}

We also plot the band structures presented by Eq. (\ref{energy band}\textbf{%
) }for several typical points in different phases and at the phase boundary
in Fig. \ref{fig4}. We note that the zero points exhibit different
geometries for different regions on the boundary. At boundaries between (I)
and (III), there are always two zero points except for the triple points. In
contrast, at boundaries between (II) and (I) or (III), there is always a
single point. In the following section, we investigate this feature from the
perspective of topology.

\section{Topological feature of zero points}

\label{Topological feature of zero points}

Now we focus on the state at the boundary of the system. From the analysis
above, we know that there are two types of boundaries, which separate two
quantum phases (I) and (III), (I) and (II) or (III) respectively. What makes
this interesting is that for the first type of boundary, the winding number
difference between two neighboring phases is $2$, while is $1$\ for the
second type of boundary. This can be seen from the equation of the loop
presented in Eq.\textbf{\ }(\ref{loop}). In general, the curve is an ellipse
with center located at $(t,0)$ in the $xy$ plane. For the first type of
boundary, the curve reduces to%
\begin{equation}
\left\{
\begin{array}{l}
x\mathbf{=}2v\cos k+t, \\
y\mathbf{=}0,%
\end{array}%
\right.
\end{equation}%
which presents a segment of $x$ axis with length $4\left\vert v\right\vert $%
\ and the center located at $(t,0)$. Along the boundary, only the length of
the segment varies. For the second type of boundary, the curve become

\begin{equation}
\frac{\left( x-t\right) ^{2}}{\left( w+v\right) ^{2}}+\frac{y^{2}}{\left(
w-v\right) ^{2}}\mathbf{=}1,
\end{equation}%
which presents a normal ellipse passing through the origin $(0,0)$. Along
the boundary, only the length of the semiaxis of the ellipse changes. To
demonstrate this point, we plot graphs presented by Eq. (\ref{loop}) to
schematically illustrate the difference between these two types of
boundaries with trivial and non-trivial topologies. It indicates that these
two types of boundaries correspond to two completely different types of
graphs. For the trivial one, the loop passes through the origin one time,
while twice times for the non-trivial one, since the loop reduces to a
segment.

\begin{figure}[tbp]
\includegraphics[ bb=45 430 371 756, width=0.45\textwidth, clip]{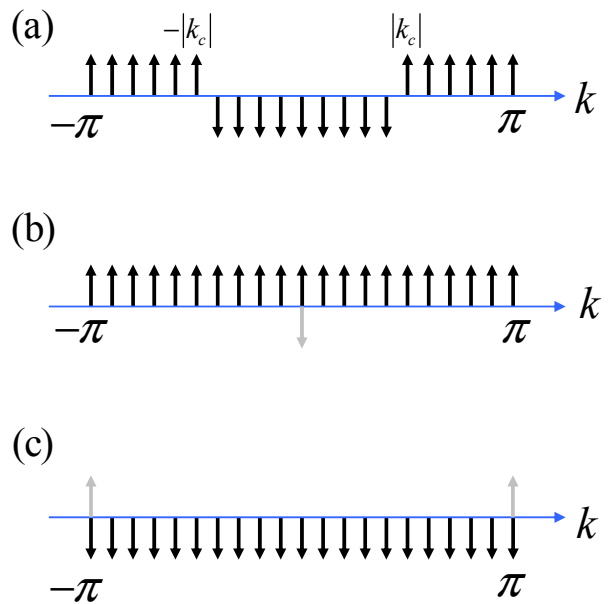}
\caption{(Color online) Schematics of topologically nontrivial field
configurations. $\pm k_c$ denote the position of band degenerate points,
which correspond to zero filled points. (a) System with $0<\vert k_c\vert<\protect\pi$. There are two kinks at $\pm k_c$ with opposite topological
charges. Two kinks can move towards or away from each other as parameters
vary, but cannot be removed until the case (b) $k_c =0$, or (c) $\vert
k_c\vert=\protect\pi$. In (b) and (c), the flip of the vectors between black
and gray arrows indicates the change of topology of the band degenerate
point. } \label{fig6}
\end{figure}

In the following, we investigate the phase boundary from the other
perspective. We introduce a $3$-D vector field $\mathbf{F}(k)$\ in $k$
space, which is defined as

\begin{equation}
\mathbf{F}(k)=(\left\langle \sigma _{x}\right\rangle _{k},\left\langle
\sigma _{y}\right\rangle _{k},\left\langle \sigma _{z}\right\rangle _{k})
\end{equation}%
where $\left\langle \sigma _{\alpha }\right\rangle _{k}=\left\langle
k\right\vert \sigma _{\alpha }\left\vert k\right\rangle $, $(\alpha =x,y,z)$%
, the expectation value of $\sigma _{\alpha }$\ for eigenstate $\left\vert
k\right\rangle $ of $h_{k}$. In the case of $v=w$, we directly obtain%
\begin{equation}
\mathbf{F}(k)=(\mathrm{sgn}(2v\cos k+t),0,0)  \label{F(k)}
\end{equation}%
where \textrm{sgn}(.) denotes the sign function. We note that the field $%
\mathbf{F}$ has a kink at the point $k_{c}=\cos [t/(2v)]$, which is the gap
closing point of the first type of the boundary. The topological charge of
the kinks are $1$ and $-1$, respectively. In Fig. \ref{fig6}, we sketch the
field $\mathbf{F}(k)$\ to illustrate the kinks. Therefore, we conclude that
the first type of nodal point is topologically protected. On the other hand,
for second type of boundary with parameters satisfying $\left\vert \left(
w+v\right) /t\right\vert $ $=1$, we have

\begin{equation}
\mathbf{F}(k)=(\mathrm{sgn}\left[ t\pm \left( w+v\right) \right] ,0,0),
\end{equation}%
which has no kink in the $k$ space. In this case, the nodal point is
topological trivial.

\section{Perturbations}

\label{Perturbations}

\begin{figure}[tbp]
\includegraphics[ bb=24 438 366 784, width=0.45\textwidth, clip]{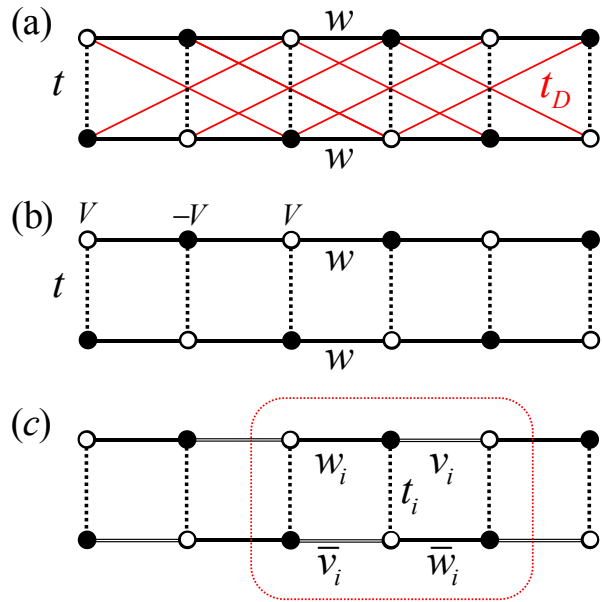}
\caption{(Color online) Schematics of two kinds of perturbation on the
coupled SSH chain at semi-metal point. (a) Diagonal hopping term (red line)
with amplitude $t_{D}$ across two neighboring plaquettes. (b) Staggered
on-site potentials on two sublattices indicated by filled and empty circles,
respectively. In the case of (a), the topology of zero-energy points is
invariant under the perturbations, while the gap opens for non-zero $V$ in
the case (b). (c) Schematic illustration of the mechanism of degenerate
point which is irrelevant to the symmetry of the system. The ladder system
is consisted of double plaquette (circled by the red dotted line). The existence
of zero-energy degenerate point requires the balance of the local parameters
$\{w_{i},v_{i},$ $\overline{w}_{i},\overline{v}_{i},$ $t_{i}\}$, presented
in Eq. (\protect\ref{parameter}). } \label{fig7}
\end{figure}

The topological invariant for the topological boundary is essentially the
topologically unavoidable\ band touching points. It seems that there is
nothing we can do to the system to get rid of the band touching point under
the restriction in Eq. (\ref{boundary1}). Changing $w$\ can only change the
location of the band touching point.\textbf{\ }This topological feature may
be robust for some kind of perturbation but fragile for others kinds. We
consider two kinds of perturbations, with an extra diagonal hopping across
two neighboring plaquettes and staggered on-site potential, respectively.
Fig. \ref{fig7} sketches the structures of two cases. We will focus on the
effects of the extra terms on the topology of the boundary.

For the first case, the Hamiltonian can be written as
\begin{equation}
H_{\mathrm{D}}=H+t_{\mathrm{D}}\sum_{j=1}^{N}(a_{j}^{\dagger }b_{j+2}+%
\mathrm{H.c.})\mathrm{,}
\end{equation}%
where $t_{\mathrm{D}}$\ denotes the diagonal hopping amplitude. Based on the
Fourier transformations in Eq. (\ref{Fourier}), we still have%
\begin{equation}
H_{\mathrm{D}}=\sum_{k}(a_{k}^{\dagger },b_{k}^{\dagger })h_{k}^{\mathrm{D}%
}\left(
\begin{array}{c}
a_{k} \\
b_{k}%
\end{array}%
\right) ,
\end{equation}%
and%
\begin{equation}
h_{k}^{\mathrm{D}}=h_{k}+2t_{\mathrm{D}}\cos 2k\left(
\begin{array}{cc}
0 & 1 \\
1 & 0%
\end{array}%
\right) .
\end{equation}%
The spectrum is%
\begin{equation}
\varepsilon _{k}^{\mathrm{D}}=\pm \sqrt{\left\vert we^{-ik}+ve^{ik}+t+2t_{%
\mathrm{D}}\cos \left( 2k\right) \right\vert ^{2}},
\end{equation}%
which only has a shift on $t$, i.e., $t\rightarrow t+2t_{\mathrm{D}}\cos
\left( 2k\right) $, from the spectrum $\varepsilon _{k}$ in Eq. (\ref{energy
band}). Thus the zero point can be obtained directly as following. Here we
consider the case that the diagonal term is added as a perturbation, i.e.,
small $t_{\mathrm{D}}$ with $\left\vert t_{\mathrm{D}}/v\right\vert \ll 1$.\
For $\sin k_{c}\neq 0,$\ $\varepsilon _{k}^{\mathrm{D}}=0$\ leads to $w=v$
and
\begin{equation}
\cos k_{c}\approx -\frac{t}{2v}\left( 1+\frac{t_{\mathrm{D}}t}{v^{2}}\right)
,
\end{equation}%
which identifies the boundary line%
\begin{equation}
w=v\text{, }\left\vert \frac{t}{2v}\left( 1+\frac{t_{\mathrm{D}}t}{v^{2}}%
\right) \right\vert <1.
\end{equation}%
On the other hand, for $\sin k_{c}=0$, the boundary is the line%
\begin{equation}
\left\vert w+v\right\vert =\left\vert t+t_{\mathrm{D}}\right\vert .
\end{equation}%
We note that for small $t_{\mathrm{D}}$, the phase diagram changes a little
comparing to the case with zero $t_{\mathrm{D}}$, but keeping the original
geometry. The position of the kink, $k_{c}$ shifts a little, without
changing the original topology, i.e., the topological charge of the kink.
Then the gapless state is topologically invariant under the perturbation
from the $t_{\mathrm{D}}-$term.

For the second case, the Hamiltonian can be written as%
\begin{equation}
H_{\mathrm{V}}=H+V\sum_{j=1}^{N}(b_{j}^{\dagger }b_{j}-a_{j}^{\dagger }a_{j})%
\mathrm{,}
\end{equation}%
which indicates that particles on different sub-lattices have opposite
chemical potentials. The extra potentials do not break the translational
symmetry.\ By the same procedure, we have%
\begin{equation}
H_{\mathrm{V}}=\sum_{k}(a_{k}^{\dagger },b_{k}^{\dagger })h_{k}^{\mathrm{v}%
}\left(
\begin{array}{c}
a_{k} \\
b_{k}%
\end{array}%
\right) ,
\end{equation}%
and%
\begin{equation}
h_{k}^{\mathrm{v}}=\left(
\begin{array}{cc}
-V & we^{ik}+ve^{-ik}+t \\
we^{-ik}+ve^{ik}+t & V%
\end{array}%
\right) .
\end{equation}%
The spectrum is%
\begin{equation}
\varepsilon _{k}^{\mathrm{v}}=\pm \sqrt{\left\vert
we^{-ik}+ve^{ik}+t\right\vert ^{2}+V^{2}},
\end{equation}%
which clearly shows that the nonzero $V$\ can open a gap, destroying the\
topological gapless state. In the vicinity of $k_{c}$, we have%
\begin{equation}
\left( \frac{\partial \left\vert \varepsilon _{k}^{\mathrm{v}}\right\vert }{%
\partial V}\right) _{k\approx k_{c}}\approx \frac{V}{\left\vert V\right\vert
},
\end{equation}%
which is discontinuous at the point $V=0$. It is a critical point of
second-order quantum phase transition, which is associated with inversion
symmetry (exchange $A$ and $B$) breaking. In fact, condition $w=v$ takes the
role of maintaining the inversion symmetry.

\section{Symmetry protection of kinks}

\label{Symmetry protection of kinks}

The kinks, as a topological invariant, should be protected by the certain
symmetry of the system. In the following, we investigate the symmetry
associated with the topological nodal points. The Hamiltonian possesses many
symmetries, for instance, translational symmetry, time reversal symmetry,
etc.. For $w=v$, the system reduces to a ladder with identical legs. There
are two specific symmetries related to the translational operator $\hat{T}%
_{1}$\ and inversion\ operator $\hat{P}$, which are defined as%
\begin{equation}
\hat{T}_{1}a_{j}\hat{T}_{1}^{-1}=b_{j+1}\text{, }\hat{T}_{1}b_{j}\hat{T}%
_{1}^{-1}=a_{j+1},
\end{equation}%
and%
\begin{equation}
\hat{P}a_{j}\hat{P}^{-1}=b_{j}\text{, }\hat{P}b_{j}\hat{P}^{-1}=a_{j}.
\end{equation}%
The action of $\hat{T}_{1}$\ shifts a lattice space along the legs of the
ladder in Fig. \ref{fig1}(a). In general, $\hat{T}_{1}$\ does not commute
with the Hamiltonian with $w\neq v$ but $\hat{T}_{1}^{2}$\ does. In
parallel, the action of $\hat{P}$\ exchange two sublattices $A$ and $B$, or
two legs of the ladder in Fig. \ref{fig1}(a). In general, $\hat{P}$\ does
not commute with the Hamiltonian except for the case $w=v$. Then in the case
of $w=v$,\ we have%
\begin{equation}
\lbrack H,\hat{T}_{1}]=[H,\hat{P}]=0.
\end{equation}%
Operators $H$,\ $\hat{T}_{1}$,\ and $\hat{P}$\ have common eigenstate set,
which can be obtained as%
\begin{equation}
\left\vert \psi _{k}^{\pm }\right\rangle =\frac{1}{\sqrt{2N}}%
\sum_{j}e^{-ikj}\left( \left\vert j\right\rangle _{b}\pm \left\vert
j\right\rangle _{a}\right) ,
\end{equation}%
with the eigen energy%
\begin{equation}
\varepsilon _{k}^{\pm }=\pm \left\vert 2v\cos k+t\right\vert
\end{equation}%
in the single-particle invariant subspace spanned by basis $\left\vert
j\right\rangle _{a}=a_{j}^{\dag }\left\vert 0\right\rangle $ and $\left\vert
j\right\rangle _{b}=b_{j}^{\dag }\left\vert 0\right\rangle $. In general,
these eigenstates satisfy%
\begin{equation}
\hat{T}_{1}\left\vert \psi _{k}^{\pm }\right\rangle =\pm e^{ik}\left\vert
\psi _{k}^{\pm }\right\rangle ,
\end{equation}%
and

\begin{equation}
\hat{P}\left\vert \psi _{k}^{\pm }\right\rangle =\pm \left\vert \psi
_{k}^{\pm }\right\rangle .
\end{equation}%
Nevertheless, the two-fold zero-energy degenerate states (with the wave
vector $k_{c}$) of $H$ can be expressed as%
\begin{equation}
\left\vert \psi _{k_{c}}^{a,b}\right\rangle =\frac{1}{\sqrt{N}}%
\sum_{j}e^{-ik_{c}j}\left\vert j\right\rangle _{a,b},
\end{equation}%
which satisfy%
\begin{equation}
\hat{T}_{1}\left\vert \psi _{k_{c}}^{a}\right\rangle =e^{ik_{c}}\left\vert
\psi _{k_{c}}^{b}\right\rangle \text{, }\hat{T}_{1}\left\vert \psi
_{k_{c}}^{b}\right\rangle =e^{ik_{c}}\left\vert \psi
_{k_{c}}^{a}\right\rangle ,
\end{equation}%
and

\begin{equation}
\hat{P}\left\vert \psi _{k_{c}}^{a}\right\rangle =\left\vert \psi
_{k_{c}}^{b}\right\rangle \text{, }\hat{P}\left\vert \psi
_{k_{c}}^{b}\right\rangle =\left\vert \psi _{k_{c}}^{a}\right\rangle .
\end{equation}%
This indicates that the nodal points at $k_{c}$\ are protected by the
symmetry related to $\hat{T}_{1}$\ or $\hat{P}$.

\begin{figure}[tbp]
\includegraphics[ bb=41 586 487 710, width=0.45\textwidth, clip]{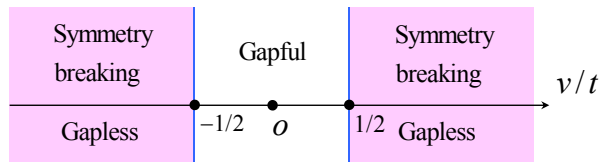}
\caption{(Color online) Schematic phase diagram of ground state of
half-filled spinless fermionic system along the axis of $w=v$. In the region
of $\vert v/t\vert>1$, the translational $T_1$ and inversion $P$ symmetries
are spontaneously broken, associated with gapless ground state. In the
region of $\vert v/t\vert<1$, the ground state is gapful and possesses the
translational $T_1$ and inversion $P$ symmetries. } \label{fig8}
\end{figure}

Furthermore, when we consider the half-filled case, the topological nodal
point is connected to a QPT associated with spontaneously symmetry breaking.
We consider the phase diagram along the axis $w=v$, which is sketched in
Fig. \ref{fig8}. A second order QPT occurs at $v=\pm t/2$,\ since the first
order derivative of groundstate energy with the respect to $v$ is
nonanalytic at the two points. This can be seen from the fact that $%
\varepsilon _{k}^{-}$\ is nonanalytic at point $k=k_{c}$. In the region of $%
|v/t|<1/2$, the gapped ground state can be expressed as%
\begin{equation}
\left\vert G\right\rangle =\prod\limits_{k}\left\vert \psi
_{k}^{-}\right\rangle
\end{equation}%
which is singlet. In the region of $|v/t|>1/2$, the ground states are
gapless and have the form%
\begin{equation}
\left\vert G_{a,b}\right\rangle =\left\vert \psi _{k_{c}}^{a,b}\right\rangle
\prod\limits_{k\notin k_{c}}\left\vert \psi _{k}^{-}\right\rangle
\end{equation}%
which are double degenerate. Since%
\begin{equation}
\hat{T}_{1}\left\vert G_{a}\right\rangle =-\left\vert G_{a}\right\rangle
\text{, }\hat{T}_{1}\left\vert G_{b}\right\rangle =-\left\vert
G_{a}\right\rangle ,
\end{equation}%
and

\begin{equation}
\hat{P}\left\vert \psi _{k_{c}}^{a}\right\rangle =-\left\vert \psi
_{k_{c}}^{b}\right\rangle \text{, }\hat{P}\left\vert \psi
_{k_{c}}^{b}\right\rangle =-\left\vert \psi _{k_{c}}^{a}\right\rangle ,
\end{equation}%
the translational $\hat{T}_{1}$ and inversion $\hat{P}$ symmetries are
spontaneously broken. Remarkably, the doubly degenerate ground states are
topologically protected by the symmetries of translation $\hat{T}_{1}$\ and
inversion $\hat{P}$. Thus, we come to the conclusion that this model
provides an example to connect the topological QPT and the one associated
with spontaneous breaking, which has been studied previously in the quantum
spin system \cite{Gang}.

On the other hand, in the case of $\left\vert (w+v)/t\right\vert =1$, the
two-fold zero-energy degenerate states are the form%
\begin{equation}
\left\vert \psi _{\pm }^{a,b}\right\rangle =\frac{1}{\sqrt{N}}\sum_{j}(\pm
1)^{j}\left\vert j\right\rangle _{a,b},
\end{equation}%
which satisfy%
\begin{equation}
\hat{T}_{1}\left\vert \psi _{\pm }^{a}\right\rangle =\pm \left\vert \psi
_{\pm }^{b}\right\rangle \text{, }\hat{T}_{1}\left\vert \psi _{\pm
}^{b}\right\rangle =\pm \left\vert \psi _{\pm }^{a}\right\rangle ,
\end{equation}%
and

\begin{equation}
\hat{P}\left\vert \psi _{\pm }^{a}\right\rangle =\left\vert \psi _{\pm
}^{b}\right\rangle .
\end{equation}%
Although $\left\vert \psi _{\pm }^{a,b}\right\rangle $ and $\left\vert \psi
_{k_{c}}^{a,b}\right\rangle $\ have the similar properties, in this
situation, we have
\begin{equation}
\lbrack H,\hat{T}_{1}]\neq 0,[H,\hat{P}]\neq 0.
\end{equation}%
The nodal point is irrelevant to the symmetry of the system. It may be due
to the mechanism of this type of degeneracy. We note that the key point of
the degeneracy arises from the balance of local parameters. As illustrated
in Fig. \ref{fig7}(c), a general ladder is consisted of many double
plaquettes. The existence of degeneracy requires%
\begin{equation}
\left\vert w_{i}+v_{i}\right\vert =\left\vert \overline{w}_{i}+\overline{v}%
_{i}\right\vert =\left\vert t_{i}\right\vert  \label{parameter}
\end{equation}%
for every double plaquette indexed by $i$. There should be infinite sets of
parameters $\{w_{i},v_{i},\overline{w}_{i},\overline{v}_{i},t_{i}\}$\
satisfy the equations. In this sense, such degeneracy is accident and
irrelevant to the symmetry of the system.

We close this section by noting that the operator $\hat{T}_{1}$\ and $\hat{P}
$\ are not antiunitary operators. This differs from the one appears in
higher dimensional system \cite{Hou}. It turns out that the time-reversal
symmetry takes a central role contributing to the protection of degeneracies
in $2$-D systems. In contrast, the model we study is a quasi-$1$-D and
possesses time-reversal symmetry, referring to nonmagnetic material.

\section{Summary}

\label{Summary}

In summary, we studied two coupled SSH chains system, which possesses a
little more complicated phase diagram in comparison with that of SSH chain.
It contains three phases, two non-trivial topological insulating phases and
one trivial phase. Moreover, the boundary between these two quantum phases
is still topologically non-trivial, which arises from two topologically
unavoidable band closing points. As inter-chain coupling strength varies,
two topological nodal points appear, move, merge and disappear in $k$ space.
We also show that the topological invariant of the boundary is protected by
the translational and inversion symmetries, rather than the antiunitary
operation in the case of $2$-D\ semimetal phase. Based on the results, we
established the possible connection between the second order QPT, associated
with a gap closing and spontaneously symmetry breaking, and the topological
QPT characterized by the change of topological invariant. Our finding
extends the understanding of topological gapless phase and provides a
platform to experimentally work within the simple system, which has
particular advantages: a lower dimension and without the need of magnetic
flux.

\acknowledgments We acknowledge the support of CNSF (Grant No. 11374163).

\end{document}